\documentstyle[aps,preprint,tighten]{revtex}
\begin{document}
\draft
\title{Sensitivity of nucleon--nucleus scattering to the off--shell
       behavior of on--shell equivalent NN potentials}
\author{H. F. Arellano$^{a}$, F. A. Brieva$^{a}$,
M. Sander$^{b}$ and H. V. von Geramb$^{b}$}
\address{$^{a}$ Departamento de F\'{\i}sica,
        Facultad de Ciencias F\'{\i}sicas y Matem\'aticas \\
        Universidad de Chile, Casilla 487--3, Santiago, Chile}
\address{$^{b}$ Theoretische Kernphysik, Universit\"at Hamburg \\
           Luruper Chaussee 149, D--22761, Hamburg, Germany}
%
\maketitle

\begin{abstract}
The sensitivity of nucleon--nucleus elastic scattering to the
off--shell behavior of realistic nucleon--nucleon interactions
is investigated when on--shell equivalent nucleon--nucleon
potentials are used.
The study is based on applications of the full--folding optical
model potential for an explicit treatment of the off--shell
behavior of the nucleon--nucleon effective interaction.
Applications were made at beam energies between 40 and 500 MeV
for proton scattering from $^{40}$Ca and $^{208}$Pb.
We use the momentum--dependent Paris potential and its local
on--shell equivalent as obtained with the Gelfand--Levitan and
Marchenko inversion formalism for the two nucleon Schr\"odinger
equation.
Full--folding calculations for nucleon--nucleus scattering show 
small fluctuations in the corresponding observables.
This implies that off--shell features of the NN interaction cannot 
be unambiguously identified with these processes.
Inversion potentials were also constructed directly from
NN phase--shift data (SM94) in the 0--1.3 GeV energy range. 
Their use in proton--nucleus scattering above 200 MeV provide 
a superior description of the observables relative to those 
obtained from current realistic NN potentials. 
Limitations and scope of our findings are presented and discussed.
\end{abstract}


\section{Introduction}
Theoretical studies of the two--nucleon interaction in their
off--shell domain have a long standing tradition
\cite{Sri75} and this topic is of renewed  interest for
designated experiments at several accelerators laboratories.
Few-- and many--body systems offer generally the possibilities
for such studies but not seldom have such endeavors ended
prematurely due to lack of statistics in data taking or an
incomplete and inconclusive theory.
Current interest on this issue comes from theory groups who have
independently developed NN  potentials which account
reasonable well for the two--body phase--shifts at energies below
pion production threshold \cite{Lac80,Mac87,Sto94}.
These potentials are manifestly different in aspects
such as their off--shell behavior and the device of an
experimental discrimination among them would be very valuable
for a comprehensive understanding
of particularities for each of the currently acceptable models.

In the past much hope was given to NN bremsstrahlung since this
three body reaction is within a Born approximation
theoretically well defined \cite{Jettergeramb,Jet95}. 
As a result of this simple reaction mechanism one  obtains a good 
perspective of the link between the $(NN\gamma)$ observables and 
half--off--shell NN {\em t} matrices. 
Using the available theoretical
developments in the field of bremsstrahlung it is now known
how to explain the available data  with different NN potentials 
as far as they reproduce the NN on--shell data very well.
This result has been independently validated and remains a surprise
in view of the obvious differences in the
half--off--shell {\em t} matrices from different model potentials.
In other words, the expectation that on--shell equivalent NN
potentials would provide distinctive bremsstrahlung predictions
was disillusioned with more complete and reliable calculations.

It can be argued that bremsstrahlung involves only
half--off--shell $t$ matrices and thus this reaction is rather
confined in phase--space.
Microscopic models of nucleon--nucleus (NA) scattering in a
full--folding framework\cite{Are90,Are94} do not suffer from such 
limitations as the optical potential depends explicitly upon the 
effective interaction fully off--shell and therefore constitute a 
wider frame for investigating the NN interaction off--shell.

Recently, significant advances have been made in accurately
handling the off--shell degrees of freedom in NA elastic scattering
\cite{Are90,Are94,Are95,Els90,Cre90,Ray92}.
Here, studies  have demonstrated that an accurate
treatment of the off--shell behavior of the NN interaction is
needed for a proper account of the theory.
Irrespective of obvious improvements in describing the data at 
projectile energies below 400 MeV, the calculations
still show some  systematic deficiencies. 
In particular the misfit of spin observables at small
momentum transfers, $\leq 1\,$fm$^{-1}$, is not  understood.
The origin of such discrepancies with the data could be attributed 
to intrinsic limitations of the bare NN potential with respect to 
the phenomenology (especially at the higher energies), to the 
simplifications implicit in the model for the NN effective 
interaction, or to the fact that the optical model potential has 
only been developed to its lowest order. 
Above 400 MeV, full folding model results quickly deteriorate as 
they fail to properly describe the observables\cite{Are90}.

Despite the significant advances in full--folding model calculations 
it remains difficult to identify,
 at the level of the scattering observables,
distinctive off--shell features of NN interactions. 
The difficulty is mainly attributed to non negligible differences 
on--shell among currently available realistic NN potentials 
\cite{Are91a}.

In this paper we investigate the sensitivity of NA scattering when 
using bare NN potentials which are equivalent on--shell.
For that purpose we have chosen the Paris potential  ($i.$) in its 
full momentum dependent form and ($ii.$) a local equivalent potential 
generated with Gelfand--Levitan--Marchenko inversion from the Paris 
potential phase--shifts and deuteron bound state.
The inversion potential yields different off--shell continuations 
as the original Paris potential. 
These differences and associated correlations  are
investigated in the full--folding  NA optical model.

We find for these two alternatives differences of about 5\% 
overall in NA observables and this rises  doubts that this  
sensitivity could be used to infer off--shell peculiarities 
of NN interactions in general. 
Based on this result we conclude that quantum inversion  provides a
practical and accurate connection between NN and NA elastic 
scattering. 
As mediating formalism may Gelfand--Levitan--Marchenko quantum 
inversion  use only {\em experimental NN data} as input and thus 
link model--independent two--body with many--body data.

This article is organized as follows.
In section II we outline the theoretical background for the
present work. We introduce and discuss aspects and assumptions
implicit in the Gelfand--Levitan and Marchenko inversion method as
well as the {\em in--medium} full--folding model of the optical
potential for NA elastic scattering.
In section III we discuss the sensitivity of NA scattering by 
calculating the scattering observables from the Paris potential and 
from an inversion NN potential constructed from the Paris NN 
phase--shifts. In this way, our analysis becomes explicitly dependent 
only from the off--shell differences between the two potentials.
In section IV we construct inversion NN potentials directly from 
the NN data, including approximately the NN phase--shifts above pion 
production threshold as a way to guide the intermediate energy 
properties of the potential.
Calculations for proton--nucleus elastic scattering in the 
40--500 MeV range and from different targets are also discussed.
Finally, in section V we present a summary and draw conclusions
from our work.

\section{Theoretical Background}

\subsection{Two Nucleon Potentials from quantum inversion}

The starting point for the NN interaction is generally
a relativistic scattering description and it is beyond any doubt,
that a realistic potential must show strong angular momentum
and spin dependencies and thus channel dependence.
An explicit momentum dependence or nonlocality is predicted
in all relativistic potential models.
Allowance for the relativistic nature of the problem in these
models is in the best cases limited to relativistic kinematics
and simplified  calculation of selected exchange diagrams.
The full dynamics of potentials requires a derivation using
quantum field theory.
This not only ensures Lorentz covariance but shows the way how
the actual potential must be constructed.
For this purpose, different formalism have been proposed and
they determine the off--energy shell or off--mass shell
amplitudes. 
Usually the derivation of such relativistic equations also
determines the propagators  and the potential which
contains all information about the dynamics of the interaction.
An accepted four--dimensional formulation of quantum field theory 
is the covariant Bethe--Salpeter equation and the three--dimensional
relativistic equations are obtained on the basis of quasi--potential
methods.
Another field theoretical methods use a generalization of the
Schr\"odinger equation in the form of the Tomonaga--Schwinger
equation and in the form of a covariant Hamiltonian formulation.
Irrespective of their foundation and motivation, the field 
theoretical models are using more or less phenomenology to fit data.

Well fitted  and widely applied  within these models
are the boson exchange potentials.
Representative within the boson exchange models are the Nijmegen
\cite{Nag75} and Bonn potentials \cite{Mac87}.
The Paris potential also belongs to this category and describes 
the long-- and  medium--range interaction with single and 
correlated two--pion exchange and  heavier meson exchanges 
\cite{Lac80}.
It fits the experimental two nucleon data reasonably well
and has tradition in calculations of
microscopic optical model potentials for NA scattering at low and
medium energies. It is generally used in its Yukawa
parameterization including  an explicit momentum dependent term
\cite{Lac80}
\begin{equation}
V = V_a+{\hbar^2\over m} p^2 V_b + V_b {\hbar^2\over m} p^2 \: ,
\end{equation}
where $m$ is the nucleon mass and $p$ the relative momentum operator.
This momentum dependence emulates a hard core repulsion
and produces a divergence of  the phase--shift
$\delta(k)$ as $k \rightarrow \infty$. 
When compared with data, the Paris potential fit quite well the 
phase shift at energies up to 280 MeV. 
Above this energy we notice a rapidly increasing divergence.
This characteristic strong repulsion at short distances
is easy to identify in cross section and spin observables
of NA scattering and  reactions\cite{Rik84}. 
Despite this weakness we have chosen the Paris potential since we 
have confidence in the numerics of our Lippmann--Schwinger and  
Bethe--Goldstone calculations  with it.

With this investigation we aim to disclose effects which are
caused by the momentum dependence and high energy phase shift
discrepancies of the genuine Paris potential in
off--shell $t-$ and $g-$matrices. 
To this purpose we use the recently developed 
Gelfand--Levitan--Marchenko inversion of partial wave radial 
Schr\"odinger equations to generate phase equivalent local 
potentials to Paris potential phase shifts. 
These inversion algorithms distinguish inversion for single and
coupled channels cases with and without a Coulomb reference
potential \cite{KohlGer94}. 
In other words, we obtain separately the hadronic part of the NN 
interaction from any set of {\em np} and {\em pp} phase shifts.
To investigate the effect of different on--shell
behaviors we also generate potentials from the latest experimental
phase shift analysis SM94 by Arndt and collaborators \cite{Arn94}.

The strong interaction inversion  potentials
$V(r;\,LSJ,T;\,np)$ and $V(r;\,LSJ,T;\,pp)$ are
numerical solutions of Gelfand--Levitan or Marchenko
integral equations,
\begin{equation}
K(r,r')+F(r,r')+\int K(r,s)F(s,r')\,ds=0
\end{equation}
and
\begin{equation}
V(r)=\pm{d\over dr} K(r,r),
\end{equation}
for any specified radius
and they are determined channel by channel.
The input kernel $F(x,y)$ is computed with the spectral information,
Jost functions or S-matrices including deuteron binding energy and
normalization constants.
We use quantum inversion as  transformation of given real
phase shifts, which are  specified within a finite energy interval
\begin{equation}
\delta(k)\sim\delta(E)=\left\{\delta(k)|E=[0,E_{max}]|
k=[0,k_{max}]\right\}  .
\end{equation}
Thereafter they
are smoothly extrapolated  with a rational function
which decays asymptotically\cite{Kir89}
\begin{equation}
\label{delta2}
\delta(E)\sim\delta(k)=\left\{\delta(k)|k\geq k_{max}|\lim
_{k\to \infty}\delta(k)\sim k^{-1}\right\} \,\,.
\end{equation}
This transformation is unique for the class of potentials in
which we are interested and which we assume physically
significant. The resulting inversion potentials are real and local.

It is evident that this inversion procedure produces a restricted 
phase equivalent potential to the Paris potential.
The limited data input $\delta(E)$ for $0\le E\leq E_{max}$ is used 
to control the range of equivalence and the extrapolation thereafter
to control the softness of the short range core interaction.

\subsection{In-medium full--folding optical potential}
\label{FF}

The optical potential for 
NA elastic scattering can be casted as the convolution of an
antisymmetrized effective interaction with the target ground state
single particle wave functions\cite{Fet65,Fes58,KMT59,Eis89,Are95}.
In momentum space this one-body operator reads
\begin{equation}
\label{ff}
U({\vec {k}\,'} , {\vec k} ; E) = 
\int d{\vec p} \: d{\vec p\,'} 
\sum_{\alpha \leq \epsilon_F}
\phi_\alpha^{\dagger}({\vec p\:'})
\left\langle
{\vec {k}\,'} , {\vec p\:'} \:
\left | F(E+\epsilon_\alpha) \right |
{\vec k} , {\vec p} \:
\right\rangle_A \: 
\phi_\alpha ( {\vec p} ) \:
\end{equation}
where $E$ represents the energy of the incoming projectile and
$\{\phi_\alpha,\epsilon_\alpha\}$ are the target ground state
single--particle wave functions and corresponding energies.
The momenta $\vec k (\vec k\,')$ and $\vec p (\vec p\,')$ 
correspond to the initial (final) momenta of the projectile 
and target struck nucleon respectively. 
The two--nucleon interaction in Eq. (\ref{ff}) accounts for multiple
scattering of nucleons to all orders in the ladder 
approximation\cite{Fet65,Kad89}. 
Although a general expression for this matrix can formally be 
defined, its practical implementation requires the device of a 
dynamical model for the effective two--nucleon interaction in 
the nucleus.
The procedure we follow is that introduced in Ref. \cite{Are95}, 
where translational invariance properties of two--nucleon 
scattering in free space or infinite nuclear matter suggest the 
following {\em ansatz} for the two-body matrix 
\begin{equation}
\label{fappr}
\left\langle
{\vec k\,'} , {\vec p\,'}
\left | \: F(\omega) \: \right |\:
{\vec k   } , {\vec p}
\right\rangle
=
\frac{1}{(2\pi)^3} \int d{\vec R}
\: e^{i {\vec R} \cdot ({\vec Q - \vec Q'})}
\left\langle {\vec {\kappa}\,'}
\left | 
f_{_{{\frac{\vec Q + \vec Q'}{2}}}} (\omega \, ; \vec R \,) 
\right |
{\vec {\kappa}} 
\right\rangle \:.
\end{equation}
Here we have defined the initial and final two--nucleon 
center-of-mass (c.m.) momenta,
\begin{equation}
\label{kcm}
\vec Q=\vec k + \vec p \: , \quad \: \vec Q'=\vec k'+ \vec p\,' \: ,
\end{equation}
and the corresponding relative momenta by
\begin{equation}
\label{krel}
\vec\kappa = {\textstyle\frac{1}{2}} ( \vec k - \vec p \,) \: , \quad \: 
\vec\kappa' = {\textstyle\frac{1}{2}} ( \vec k'- \vec p\,')  \: .
\end{equation}
The function 
$\langle{\vec {\kappa}\,'} | f_{ \vec Q } (\omega \, ; \, \vec R \,) 
| {\vec {\kappa}} \rangle$  
corresponds to the matrix elements of a reduced two-body effective 
interaction. 
In the case of no dependence of the $f$ matrix upon the spatial 
coordinate ($\vec R$) one restores total momentum conservation of 
the interacting pair as the radial integral in Eq. (\ref{fappr})
leads to a c.m. momentum conserving Dirac $\delta-$function.

A calculable expression for the optical potential {\em in medium} 
emerges after a systematic reduction of the many body propagator 
when represented in terms of the target ground state spectral 
function\cite{Are95}. 
To lowest order in a series expansion of the two-body propagator 
in a finite nucleus this interaction can be identified with the 
$g$ matrix solution of the Brueckner-Bethe-Goldstone equation
for interacting nucleons in infinite nuclear matter evaluated at 
nuclear density $\rho(R)$ in the nucleus. 
Furthermore, it becomes convenient at this point to substitute 
the single particle energies $\epsilon_\alpha$ by an average value, 
$\overline\epsilon$, and to use the Slater or Campi-Bouyssy 
approximations\cite{Cam78,Are90b} to represent the ground state 
mixed density, i.e. 
\begin{equation}
\label{mixed}
\rho \left ( \vec p\,' , \vec p \right ) = 
\sum_{\alpha \leq \epsilon_F}
\phi_\alpha^{\dagger} ( \vec p\,' )  
\phi_\alpha           ( \vec p\,  ) 
\approx 
\frac{4}{(2\pi)^3} \int\! d{\vec R} 
\: e^{i ({\vec p\,'-\vec p \,}) \cdot{\vec R}} \rho(R) \nonumber \\
\left \{
\frac{1}{\hat\rho(R)} \int\! d{\vec P} \: \Theta[\hat k(R) - P] \:
\right \}  \: ,
\end{equation}
%
where $\rho(R)$ is the local nuclear density at coordinate $\vec R$
and $\vec P$ represents the struck nucleon mean momentum defined by
\begin{equation}
\label{p}
\vec P = {\textstyle\frac{1}{2}}(\vec p + \vec p\,') \, .
\end{equation}
The local momentum function $\hat k(R)$ sets 
the range of variation of the struck nucleon mean momenta upon
collisions with the projectile and is  obtained from the Slater 
or Campi-Bouyssy\cite{Cam78} prescriptions. 
The local density function $\hat \rho(R)$ is defined in terms of 
the local momentum function by
\begin{equation}
\label{rhohat}
\hat \rho(R) = \frac{2}{3\pi^2} \hat k^3(R) \: .
\end{equation}
With the above considerations the optical potential
can be expressed in terms of the nuclear density and a Fermi
averaged effective interaction obtained from 
interacting nuclear matter. 
This interaction retains nuclear medium correlations associated 
with the nuclear mean fields and Pauli blocking.  
The {\em in-medium} full--folding optical potential then reads
\begin{eqnarray}
\label{ffg}
U({\vec {k}\,'}, {\vec k} ; E) & = &
\frac{4}{(2\pi)^3} \int\! d{\vec R} 
\: e^{i {\vec q} \cdot {\vec R} } \rho(R)
\frac{1}{\hat\rho(R)} \int\! d{\vec P} \: 
\nonumber \\ 
& \times &
\Theta[\hat k(R) - P] \:
\left\langle
{\textstyle\frac{1}{2}}( \vec K - \vec P - \vec q )
\left | 
g_{_{\vec K + \vec P }} (E + \overline\epsilon \, ; \vec R \, ) 
\right |
{\textstyle\frac{1}{2}}( \vec K - \vec P + \vec q )
\right\rangle_A
\: .
\end{eqnarray}
Thus, the optical potential  requires the calculation 
of $g$ matrices off--shell as their relative momenta obey no 
constraints apart from those imposed by the ground state mixed 
density of the target. 
Furthermore, no assumptions are introduced on the nature of the 
momentum dependence of the optical potential, thus retaining all 
non localities arising from the genuine momentum dependence of 
the NN effective interaction and as prescribed by the full--folding 
integral.
Actual calculations involve determining $g$ matrices at several 
densities and over a wide range of total center-of-mass momenta, 
features fully accounted for in the present work.

In the context of a medium independent internucleon interaction, as
when the free $t$ matrix is used to represent the NN
effective interaction, the integral over the spatial coordinate in
Eq. (\ref{ffg}) can be performed separately from the motion of the 
target nucleons. 
With the use of Eq. (\ref{mixed}) for the mixed density one recovers 
the expression for the full--folding optical
potential in the zero density approximation\cite{Are90}, 
namely $U({\vec {k}\,'} , {\vec k} ; E) \: \rightarrow
U_{o}({\vec {k}\,'} , {\vec k} ; E) $, where
\begin{equation}
\label{fft}
U_{o}({\vec {k}\,'} , {\vec k} ; E)  =
\int d{\vec P}
\: \rho({\vec P} + {\textstyle\frac{1}{2}}{\vec q} ,
		 {\vec P} - {\textstyle\frac{1}{2}}{\vec q} )
\left\langle
{\textstyle \frac{1}{2}} (\vec K - \vec P - \vec q)
\left | t_{_{{\vec K} + {\vec P}}} (E + \overline\epsilon) \right |
{\textstyle \frac{1}{2}} (\vec K - \vec P + \vec q)
\right\rangle_A \: .
\end{equation}

The dependence of the optical potential on off--shell
$t-$matrices becomes explicit in the above expression. 
The feasibility of the full--folding model to investigate 
particular signatures of the effective interaction off--shell
will depend on the sensitivity
of NA scattering observables to the use of 
$t$ matrices with manifestly distinctive behaviors off--shell. 
An important constraint for such study is that the effective 
interactions, $t$ matrices in this limit, agree on--shell.
To the extent this constraint is met, one can attain the 
differences in the NA scattering observables to the 
differences of the interactions off--shell.

A simple kinematical effect usually overlooked, but explicitly 
accounted for in our calculations, is that the full--folding 
approach calls for matrix elements of energy $E+\overline\epsilon$ 
in the laboratory frame. 
In the limit of the free $t$ matrix for the NN interactions, 
the energy of the interacting pair in its c.m. is given by
$E+\overline\epsilon - {\textstyle\frac{1}{4m}}(\vec K + \vec P)^2$. 
Therefore, the maximum energy of the pair in its c.m. is 
$E+\overline\epsilon$, the energy of the beam plus the average
binding energy of the target nucleons. 
This is to say that in the case of optical potentials for 
nucleons at 500 MeV, $t$ matrices of up to $\sim$1 GeV in the 
laboratory frame are required. 
This more demanding sampling of the NN effective interaction is a 
result of the unconstrained kinematics allowed by the Fermi motion 
of the nucleons in the nucleus.

A few comments are pertinent regarding further approximations in 
the treatment of the $t$ matrix which limit a clear assessment of
the off--shell behavior of the NN interaction in NA scattering.
A simplifying assumption, commonly used in some alternative 
full--folding calculations\cite{Els90,Cre90}, is that the $t$ matrix 
varies very weakly with respect to the NN c.m. momentum
$\vec K + \vec P$.  
Thus, the magnitude of this momentum is fixed to the (asymptotic)
on--shell value of the incoming projectile, $K_o$, and the $t$ 
matrix is approximated by 
\begin{equation}
t_{_{\vec K + \vec P}} (\omega ) 
\approx 
t_{_{\vec K_o}} (\omega ) \;.
\end{equation}
Thus, $t$ matrices are evaluated at a fixed energy equal in the 
NN c.m.  to one--half the energy of the beam.
Here one has neglected all effects associated with the Fermi 
motion in the NN c.m. momentum dependence.
The resulting full--folding calculations samples the $t$ matrix 
off--shell through its dependence on $\vec P$ in the relative 
momenta exclusively (see Eq. (\ref{fft})).
This approximation seems adequate at beam energies near 300 MeV. 
Its application at lower or higher energies, however, needs further
considerations as the NN $t$ matrix exhibits a sizable NN c.m. 
momentum dependence\cite{Are94}.
In the low energy region, apart from the fact that medium 
effects need to be incorporated in the model, the underlying 
kinematics prescribed by the full--folding yields the sampling 
of the $t$ matrix in regions where it varies significantly as 
the low energy behavior of the interactions becomes dominant.
In the high energy regime, in turn, difficulties arise from the 
opening of inelastic channels such as those associated with pion 
production or $\Delta-$resonances. 
The actual merit of the theory of the optical potential needs to 
be assessed with a consistent incorporation of such additional 
degrees of freedom.
 
\subsection{ The NN effective interaction}

In the present approach, correlations associated with interacting 
nucleons in the nuclear medium are obtained from the NN effective 
interaction defined by the Brueckner-Bethe-Goldstone equation 
for symmetric nuclear matter. In momentum representation,
the $g$ matrix associated with interacting nucleons of total
c.m.  momentum $\vec Q$, starting energy $\omega$ and nuclear
density $\rho$ satisfies 
\begin{equation}
\label{gmatrix}
\left\langle {\vec {\kappa}\,'}
\left | g_{_{\vec Q}}(\omega \, ; \vec R \, ) \right |
{\vec {\kappa}} \right\rangle =
\left\langle {\vec {\kappa}\,'} \left | V \right |
{\vec {\kappa}} \right\rangle \:
+ \int d{\vec {\kappa}\,''} \:\left\langle {\vec {\kappa}\,'}\left |
 V \right | {\vec {\kappa}\,''}
\right\rangle \:
\lambda^{^{NM}}_{\vec Q}({\vec {\kappa}\,''} ; \omega ; k_F \, )
\left\langle  {\vec {\kappa}\,''} \left | g_{_{\vec Q}}
(\omega \, ; \vec R \, ) \right | {\vec {\kappa}}\right\rangle \:,
\end{equation}
with $k_F$ the nuclear matter Fermi momentum determined from the 
nuclear density {\em via}
\begin{equation}
\label{kf}
k_F = \left ( \frac{3\pi^2}{2}\rho (R) \, \right )^{1/3} \: .
\end{equation}
The two-body propagator $\lambda^{^{NM}}$ models both Pauli blocking
and the nuclear mean field effects in the propagation of intermediate 
states, 
\begin{equation}
\label{lambda}
\lambda^{^{NM}}_{_{{\vec Q}}}({\vec q} \: ; \: \omega \: ; \: k_F) =
\frac{ {\cal Q}( P_{+}\: ; P_{-}\: ; k_F\: )}
     {
\omega + i \eta - \epsilon(P_{+} ; k_F) -\epsilon(P_{-} ; k_F)
     } \: ,
\end{equation}
with 
$P_\pm =\left |{\textstyle\frac{1}{2}}\vec Q \pm\vec q \,\right | $
and $\cal Q$ the Pauli blocking function
\begin{equation}
\label{pauli}
{\cal Q}( P_{+}\: ; P_{-}\: ; k_F \:) =
\Theta[\: \epsilon(P_{+} ; k_F) - \epsilon_F \:] \:
\Theta[\: \epsilon(P_{-} ; k_F) - \epsilon_F \:] \: \: .
\end{equation}
Here the single particle energies $\epsilon$ are defined in terms
of the self-consistent nuclear matter fields, $U_{NM}$,
\begin{equation}
\label{epsilon}
\epsilon(k_\alpha ; k_F) = 
\frac{k_\alpha^2}{2m}+Re\left [ U_{NM}(k_\alpha ; k_F) \right ] \, ,
\end{equation}
where the mean fields $U_{_{NM}}(k;\:k_F)$ are calculated 
self--consistently for the underlying bare NN interaction from
\begin{equation}
\label{unm}
U_{_{NM}}(k;\:k_F) = \sum_{\alpha \leq {\epsilon}_F} \left\langle
{\textstyle\frac{1}{2}}({\vec k} -
{\vec k}_{\alpha} ) \left | 
g_{_{\left[\, {\vec k} + {\vec k}_{\alpha}\,\right]}}
(\epsilon (k) + \epsilon (k_{\alpha}\:)) \right |
{\textstyle\frac{1}{2}}({\vec k} - {\vec k}_{\alpha} ) 
\right\rangle \: ,
\end{equation}
Actual calculations of these mean fields have been made using the 
continuous prescription at the Fermi energy\cite{Jeu76} and 
simplifying the Pauli blocking function $\cal Q$ by its 
angle--averaged form.

\section{Paris versus Paris inversion}

We have used the Paris potential to generate a set of NN 
phase--shifts which are taken as input to calculate the 
corresponding inversion potential. 
Thus, we make sure that Paris and the inversion potential 
are equivalent on--shell within the accuracy obtained with 
the numeric algorithms used for implementing the quantum 
inversion method. 
The range of energies where the explicit phase--shifts are 
considered determine importantly the off--shell behavior of 
the inversion potential. 
This effect is observed when calculating, for example, the 
optical potential and observables for NA scattering from the 
free $t$ matrix obtained from inversion potentials constructed 
from different sets of phase--shifts. 
Therefore, to ensure that properties of the inversion potential 
depend solely on the dynamical equations and not on the range 
of energies considered for the phase--shifts, we have verified 
that a set of Paris phase--shifts in the 0--1.3 GeV energy range 
is sufficient to construct an inversion potential that can be 
used up to 500 MeV in effective interaction calculations (Sec. II.C).
Above 1.3 GeV, the inversion algorithm assumes a smooth decrease 
to zero at infinity for the phase-shifts (Eq. \ref{delta2}).
This extrapolation departs strongly from the Paris potential 
behavior.

In Fig. \ref{F1} we show the Paris phase-shifts (crosses) and the 
phase-shifts obtained from its corresponding inversion potential 
(full line) for selected NN channels ($L \leq 2$) and in the 
0--1.2 GeV energy range. 
In general, for most of the NN channels we have considered the 
agreement is excellent.
Some differences are observed in the (coupled) $^3D_1$ state 
above 400 MeV. 
Altogether, we can conclude that the calculated inversion
potential is phase--equivalent to the Paris potential. 
Their differences in a many--body system should come from the 
intrinsic properties of the two potentials and provided that 
the off--shell sampling is compatible with the energy range 
used to set the phase equivalence.

The  inversion potential differs on its off--shell content from 
the original Paris potential as the former is static and local, 
whereas the latter is momentum dependent. 
These differences and associated correlations to all
orders are now investigated in the context of the full--folding 
model of the optical potential for NA scattering and as described 
in Sec. II.B.

\subsection{Sensitivity to off--shell effects in NA scattering}

We have calculated both $t$ matrices and {\em in--medium} $g$ 
matrices from the Paris and its inversion potentials. 
The inversion scheme was used for all the NN channels with total 
angular momentum $J \leq 2$. 
For channels with $J > 2$ the genuine Paris potential is used. 
The nucleon--nucleus optical potential was calculated using the 
$g$ matrix as effective interaction since medium effects have 
been proved important even for nucleons with incident energies 
of 400 MeV \cite{Are95}. 
The corresponding $g$ matrices were calculated solving 
Eq. (\ref{gmatrix}) using standard matrix inversion 
methods\cite{Adh91}. 
The ${\vec R}-$dependence in the full--folding integral 
was obtained by calculating $g$ matrices at different densities 
as obtained from different values for $k_F$ up to 1.4 fm$^{-1}$. 
For an accurate off--shell sampling of the NN effective
interaction, $g$ matrices were calculated at several 
values of the total NN c.m. momentum in the 0 -- 7 fm$^{-1}$ 
interval, with higher density of points in the region it varies
most rapidly as a function of the c.m. momentum. 
Contributions associated with the deuteron bound state singularity
were also included\cite{Are94}.
The ground state nuclear densities and average binding energies 
($\overline\epsilon$ in Eq. (\ref{ffg})) are the same as in 
Ref. \cite{Are95}.

Calculations of differential cross sections ($d\sigma/d\Omega$) and
analyzing powers ($A_y$) were made for proton elastic scattering 
from $^{40}$Ca and $^{208}$Pb in the 40--400 MeV energy range. 
Results for the spin rotation function have been omitted for
brevity as they exhibit similar behavior as those observed in $A_y$.
In Figs. \ref{F2} and \ref{F3} we present the calculated scattering
observables, as function of the center of mass scattering angle 
($\Theta_{c.m.}$) in the 40--MeV application, and as a function of
the momentum transfer ($q$) at 400 MeV.
The data for p+$^{40}$Ca scattering data at 40 MeV were taken from 
Ref. \cite{Mcc86,Blu66}.
In the case of p+$^{208}$Pb, the data at 400 MeV were taken from 
Ref. \cite{Hut88}.
The full and dashed curves represent results for the Paris 
potential and corresponding inversion respectively. 
Results at different energies and for different targets show 
comparable quantitative differences.
Also, very similar differences in the observables are observed when 
using a free $t$ matrix as input in the full--folding calculations.

Some conclusions can be drawn. 
The two phase--equivalent potentials give a qualitatively similar 
description of the scattering observables, regardless
their very different intrinsic structure. 
The differences between the two curves in Figs. \ref{F2} 
and \ref{F3} reflect the level of sensitivity to the off--shell 
behavior associated with each underlying NN effective interaction. 
Quantitatively, we observe up to a 10\% difference in the magnitude 
of the calculated observables due to differences off-shell
in the effective interaction, feature that is maintained in 
the energy range under study. 
This weak sensitivity to differences off--shell of completely 
different NN bare potentials suggests that the full--folding 
model cannot unambiguously discriminate among on--shell equivalent 
NN potentials. 
Furthermore, this equivalence indicates that a determining element
in the NN potential is its on--shell content. 
Indeed, we have tested this by taking a restricted energy 
range (0 - 400 MeV) for the NN phase--shifts 
to construct the inversion potential. 
In this case, the off-shell content of the inversion potential is 
different as reflected by the differences observed 
in the NA scattering observables, more significantly above 200 MeV. 
We have also considered a 0--3 GeV energy range for the genuine 
Paris phase--shifts to construct the inversion potential. 
In this case we observed no further differences
with respect to those observed when using the 0--1.3 GeV energy
range.
These results indicate that NN potentials 
which closely accounts for the NN phase--shift data over a 
wide energy range one would be able to assess the level 
of completeness of the optical model for NA scattering. 

The relatively weak sensitivity to the off--shell effects suggest
that NN potentials, constructed by means of the inversion scheme and 
following closely the NN data are meaningful and would provide 
predictions for the full--folding 
model very close to what would it be obtained from first principle
NN potentials with comparable fit to the data.

\subsection{Effective interactions off--shell}

To illustrate the degree of sensitivity of the 
$g$ matrix upon alternative choices of bare NN potentials, 
we have calculated selected matrix elements relevant for 
the leading contribution to the optical potential.
Thus, we consider diagonal $g-$matrix elements, i.e.  
$\langle \vec \kappa | g_{ \vec Q_o; k_F } (\omega \, ; \, \vec R \,) 
| \vec \kappa \rangle$, with $Q_o$ fixed to a single value by
$Q_o = \sqrt{2 m \omega}$.
In the context of the free $t$ matrix ($k_F = 0$) this kinematics
implies the on--shell relative momentum occurs for 
$\kappa={\textstyle\frac{1}{2}}Q_o$.
Since $\kappa$ is in general independent of both $Q_o$ 
and $\omega$, the resulting function corresponds to off--shell
elements of the $g$ matrix. 
For the Paris  and its inversion equivalent  potential we have 
solved Eq. (\ref{gmatrix}) for $g$ in the cases $k_F$=0 fm$^{-1}$ 
and $k_F$= 1 fm$^{-1}$ and for the state $^1S_0$. 
This is a good example to show the differences generally 
observed in most of the states for the effective interactions.
In Fig. \ref{F4} we show the corresponding matrix elements, both 
real and imaginary components, for $\omega$= 30, 200 and 400 MeV. 
The solid and the dashed curves represent results from the Paris 
and the Paris inversion potentials respectively.

Since the Paris potential and its inversion are equivalent on--shell, 
the solid and dashed curves for the $t$ matrix ($k_F$=0) must 
intercept each other at $\kappa={\textstyle\frac{1}{2}}Q_o$ in 
both their real and imaginary components.
This fact is indeed the case in the figures shown here as the 
on--shell constraint is explicitly built in the inversion method; 
this is not necessarily the case, however, for the $g$ matrix as 
the propagator differs from that in free space by the presence of 
Pauli blocking and the self--consistent fields.
It is interesting to note the different asymptotic behavior 
($k\to\infty$) between the Paris and the inversion potential for 
both $k_F$=0 and $k_F$=1 fm$^{-1}$.
Whereas all curves tend to zero as k increases, the one 
corresponding to the real part of the $g$ matrix for the Paris 
potential does not.
This result is consistent with the explicit momentum dependence 
introduced in the parameterization of the Paris potential, feature 
which becomes dominant at high momenta.
The extent to which this distinctive behavior is significant
in the dynamics of the collision of the projectile with
the nucleus needs to be assessed in the context of the elastic
scattering observables.

In Fig. \ref{F4} we also observe that the plotted matrix 
elements usually differ utmost in a few percents around the 
on--shell values.
This result is consistent with the differences we have already 
discussed at the level of the NA observables.

\section{NA scattering from NN phase-shifts}

In this section we extend the idea of NN potentials obtained 
directly from NN phase--shifts through the quantum inversion 
method and its application to NA scattering. 
For this purpose, we have calculated a NN inversion potential 
based on the SM94 phase--shift analysis of Arndt and collaborators 
\cite{Arn94}. 
The main problem we face is the choice of a meaningful energy 
range where the phase--shifts are to be taken from. 
Indeed, in order to be consistent with the inversion scheme, a 
set of real phase--shifts are needed. 
This sets a limit for the energy range at the pion production 
threshold ($\sim$300 MeV). 
However, our studies with the Paris potential in sect. III show 
that phase--shifts at much higher energies are required by the 
inversion method.

Our approach is as follows. 
In order to account approximately for the trend with energy that 
phase--shifts have above pion production threshold, we have 
neglected the imaginary component of the phase--shifts and 
retained only their real components. 
Thus we can construct a real NN inversion potential from a
set of real phase--shifts. 
The range of energy considered for the inversion is 0--1.3 GeV. 
Although this approach becomes essentially qualitative, we expect
that our calculations will provide a guidance on how the medium and
short range parts of the NN interaction, as determined by more 
realistic phase--shifts, affect the description of NA scattering 
in the intermediate energy region.

We have calculated NN inversion potentials from the SM94 data for 
all NN channels with $J \leq 2$. In Fig. \ref{F5} we present the
phase--shifts as a function of the energy for some selected channels. 
Dots correspond to the SM94 data (only the real part of the 
phase--shifts) and full curves are the results from the inversion 
potentials constructed from the SM94 data. 
We also plot the phase--shifts obtained from the Paris
inversion potentials as reference (dashed curves). 
These results suggest two observations. 
One is related to the ability of the inversion method to reproduce 
the input data.
The level of accuracy obtained is again excellent, as in the 
Paris case discussed in sect. III. 
The other point of physical significance is the clear departure of 
the Paris phase--shifts from, at least, the real part of the 
experimental ones above pion production threshold. 
It is precisely the presence of these differences which should 
affect the overall behavior of the inversion potential and, in 
particular, its off--shell components. 

The role of the NN inversion potentials based on the SM94 data has 
been tested in NA elastic scattering. 
We have performed {\em in--medium} ($g-$matrix) full--folding 
calculations for proton scattering on $^{40}$Ca at several energies. 
In the actual calculations, the inversion potentials were used for
the NN channels with $J \leq 2$. 
For simplicity, we have kept the components corresponding to the 
Paris potential for all the states with $J > 2$. 
In Fig. \ref{F6}  we show the differential cross--section and 
analyzing power ($A_y$) results for p + $^{40}$Ca at 40 MeV. 
The full line corresponds to the results from the inversion 
potential and the dashed curve to results from the Paris inversion 
one. 
In this and following figures we include as reference the 
calculations done with Paris inversion potentials to have a 
clear comparison of the differences that may be observed. 
In the scheme we have followed, these differences would come 
directly from the different underlying set of NN phase--shifts used. 
The results in Fig. \ref{F6} at 40 MeV show, however, little 
difference. 
This is consistent with the fact that both potentials are comparable 
in their agreement with the NN data below pion production threshold 
(see Fig. \ref{F5}). 
Still, some small sensitivity of the order of 5 \% in the observables 
is observed. 
Since both potentials are constructed following exactly the same 
inversion procedure, we conclude that phase--shifts above 300 MeV 
still determine properties of the NN potentials which affect 
low--energy NA scattering. 
These small, mostly off--shell, differences stay in even at 
much smaller energies in many--body systems. 

We have pursued these calculations at higher energies. 
In Fig. \ref{F7} we present the results for p + $^{40}$Ca at 
200 and 300 MeV. 
The meaning of the curves is the same as in Fig. \ref{F6} and 
the data was taken from Ref. \cite{Kel89} at 200 MeV and from
Ref. \cite{Ste85} at 300 MeV. 
At these energies we start noting a marked departure in the 
predictions of the two inversion potentials, with a tendency 
of the potential constructed from the NN phase--shift data to 
be relatively closer to the NA scattering data. 
This result reflects the disagreement existing between the data
and the Paris potential phase--shifts (Fig. \ref{F5}). 
Our findings are confirmed with calculations for p + $^{40}$Ca 
at 400 and 500 MeV. 
These results are shown in Fig. \ref{F8}. 
Here, the data was taken from Ref. \cite{Ble88} at 400 MeV and 
from Ref. \cite{Hut88} at 500 MeV. 
Certainly, applications of the Paris (or its inverse) potential at
400 MeV and above constitute an extrapolation of the model. 
Nevertheless, these applications serve us to illustrate  both 
the role of the NN phase--shifts above pion production threshold 
in determining the NN potential and the ability of the inversion 
method to capture that physics.
The differences given by the two potentials in Fig. \ref{F8} are 
remarkable. 
In particular, it is notable the improvement obtained in describing 
the NA scattering observables with the inversion potential 
constructed from the SM94 data, mainly for $q>$1 fm$^{-1}$ in 
both d$\sigma$/d$\Omega$ and A$_y$.
Furthermore, uncertainties associated with the off--shell behavior 
of the NN potential are smaller than the departure of the inversion 
potentials from the Paris to the SM94 data. 
This indicates that the improvement in the description of the NA
data is a direct consequence of an improved account for the NN data,
mainly above pion production threshold,
a built--in feature in the inversion potential.

\section{Summary and Conclusions}

In this paper we have addressed the problem of how properties of 
the underlying NN bare interaction determines the dynamics of a 
many--nucleon system. 
Our approach for the NN force is based on the quantum inversion 
method where a local, static, channel dependent potential is 
constructed directly from the NN phase--shifts. 
By departing from NN potentials derived from a field theoretical 
approach and empirically modified to maximize the fit to NN
data below pion production threshold, we expect to shed new light 
on these properties of the NN force relevant to many--body 
processes.

We have investigated effects associated with the genuine off--shell
behavior of bare NN potentials which are equivalent from the point
of view of the phase--shifts over a wide energy range.
The generation of on--shell--equivalent potentials 
was based on the Gelfand--Levitan and Marchenko inversion
method for the NN system.
Applications were made with the Paris potential, where the
inversion method was applied to its corresponding phase--shifts 
up to 1.3 GeV kinetic energy in the NN laboratory system.
This fairly large energy range was required to have the off--shell 
behavior of the inversion potential being determined mainly by the 
theory and not by the choice of a particular set of phase--shifts.
The corresponding NN effective interactions were found to 
exhibit sizable differences off--shell, particularly for 
relative momenta above 3 fm$^{-1}$ in the NN system. 
The investigation of these differences was made in the context 
of NA elastic scattering with the calculation of {\em in--medium}
full--folding optical potentials for proton scattering from
$^{40}$Ca and $^{208}$Pb and at beam energies between 40 
and 400 MeV. 
We found a weak sensitivity of the NA scattering observables 
to the off---shell differences  observed between the NN effective 
interactions provided the underlying bare NN potentials are 
equivalent on--shell. 
These differences, at most 10\% at the level of the NA scattering 
observables, are present in the whole energy range.
One striking aspect is the ability of the inversion method to 
generate an NN potential which is, essentially, physically 
equivalent to its original counterpart although, by construction, 
their analytic properties and asymptotic behavior are very different.

Based on the success of the inversion method, we have constructed 
inversion NN potentials based on the SM94 phase--shift analysis. 
These potentials have been applied to the calculation of 
full--folding optical potentials for proton elastic scattering 
from $^{40}$Ca in the 40--500 MeV energy range. 
We obtained results for the scattering observables which yield a 
fit to the NA scattering data at 40 MeV comparable to that 
obtained with the Paris potential. 
However, a departure from the Paris results appears above 200 MeV, 
with a clear improvement of the NA scattering data being achieved 
in the 400-- and 500--MeV applications. 
This improvement comes as a direct consequence of the closer
agreement between the inversion potential and the NN phenomenology.
An important conclusion emerging from our studies is that bare NN 
potentials which only provide a fit to NN phase--shifts below pion 
production threshold and disregard their higher energy behavior are 
unlikely to be realistic candidates to describe nucleon--nucleon 
dynamic in the nuclear medium.

Despite the improvements in the description of NA scattering using
inversion potentials from NN data, 
difficulties still remain in describing 
the NA scattering data at momentum transfers below 1 fm$^{-1}$. 
We believe the explicit treatment inelasticities of the NN 
interaction above pion production threshold and baryon excitation 
mechanisms need to be addressed as the inversion potentials 
from the SM94 analysis were restricted to the real part of 
the phase--shifts only.
On the other hand, in the lower energy applications we observe 
a systematic inability of the full--folding model to describe in 
greater detail the NA scattering data.  
Here, the presence of higher order processes due to the
many--body nature of the problem need definitely be accounted for.
Indeed, the results obtained at these energies using the Paris
potential, its inversion and the SM94 inversion provide
essentially the same description of the NA scattering data. 
Therefore, we cannot attain these limitations of the optical model
to the uncertainties associated with the off--shell behavior of 
the NN interaction.

\begin{acknowledgments}
H.F.A. is grateful for the hospitality of the nuclear theory group 
of the Theoretische Kernphysik, Universit\"at Hamburg.
This research was supported in part by FONDECYT grants 3940008 
and 1931115, and the Forschungszentrum J\"ulich 
COSY Collaboration Grant No. 41126865. 
\end{acknowledgments}

%
%
%
%

%
%
%
\begin{figure}
\caption{$L \leq 2$ channel phase--shifts from the
Paris potential (crosses) and from corresponding inversion potential 
(full curves).}
\label{F1}
\end{figure}
\begin{figure}
\caption{Scattering observables calculated from the Paris (solid curves)
 and its corresponding inversion potential (dashed curves)
 for p + $^{40}$Ca at  40 MeV.}
\label{F2}
\end{figure}
\begin{figure}
\caption{Scattering observables calculated from the Paris (solid curves)
and its corresponding inversion potential (dashed curves) for 
p + $^{208}$Pb at 400 MeV.}
\label{F3}
\end{figure}
\begin{figure}
\caption{Behavior of the diagonal elements of the $g$ matrix for
the state $^1S_0$
as a function of the relative momentum and for $\omega$=30 MeV
(top), $\omega$=200 MeV (center) and $\omega$=400 MeV (bottom).
The figures on the left correspond to $k_F = 0$ ( $t$ matrix) and
those on the right to $k_F=1.0 \,fm^{-1}$
The solid curves represent results from the genuine Paris potential.
The dashed  curves correspond to results obtained from inversion 
potentials based on the Paris phase--shifts.
}
\label{F4}
\end{figure}
\begin{figure}
\caption{$L \leq 2$ channel real phase--shifts from SM94 data (dots),
and from the inversion potentials constructed from the real part of the
SM94 data (solid curves) and the Paris potential (dashed curves).}
\label{F5}
\end{figure}
\begin{figure}
\caption{Calculated and measured differential cross-section
and analyzing power for p+$^{40}$Ca elastic scattering at 40 MeV.
The solid and dashed curves were obtained from
full--folding using the SM94  and Paris inversion,
respectively.
All curves represent {\em in-medium} full--folding calculations.}
\label{F6}
\end{figure}
\begin{figure}
\caption{Calculated and measured differential cross-section
and analyzing power for p+$^{40}$Pb elastic scattering at 200 and 300 MeV.
The curve patterns follow convention of Fig. \protect\ref{F6}.}
\label{F7}
\end{figure}
\begin{figure}
\caption{Calculated and measured differential cross-section
and analyzing power for p+$^{40}$Ca elastic scattering at 400 and 500 MeV.
The curve patterns follow convention of Fig. \protect\ref{F6}.}
\label{F8}
\end{figure}
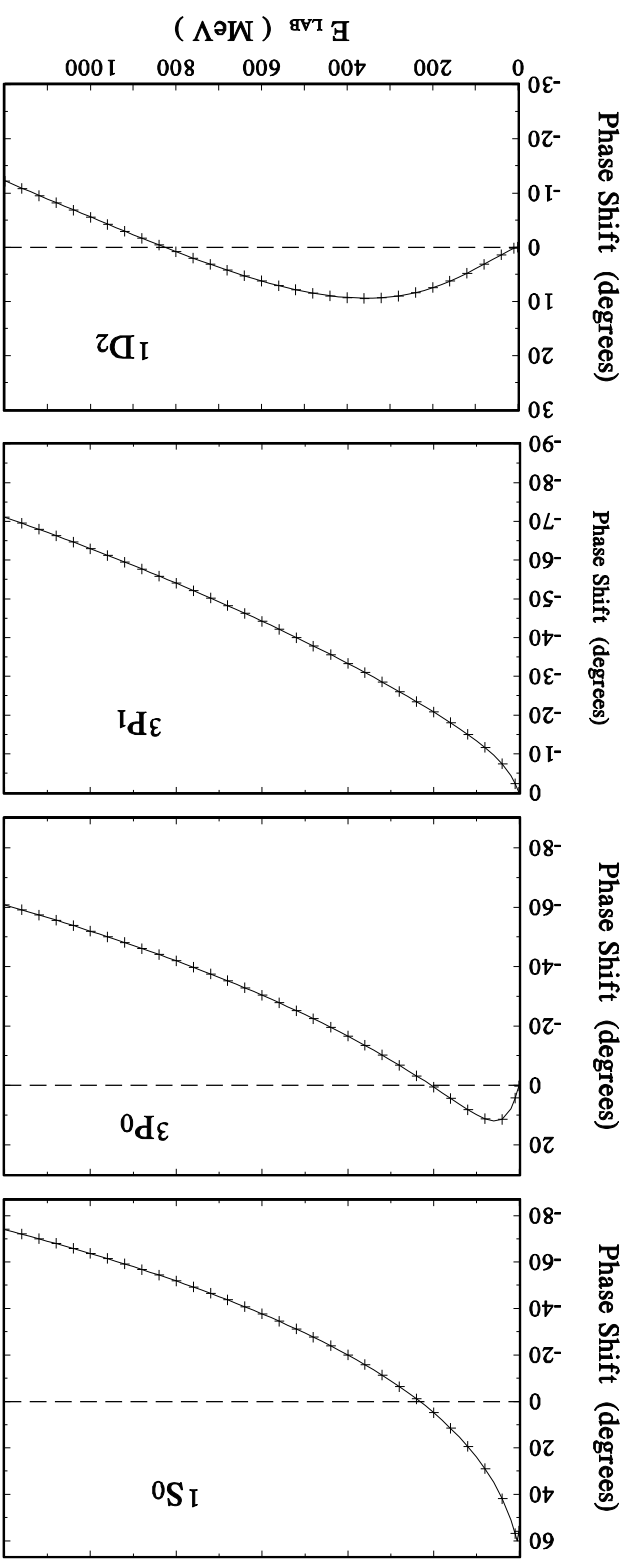
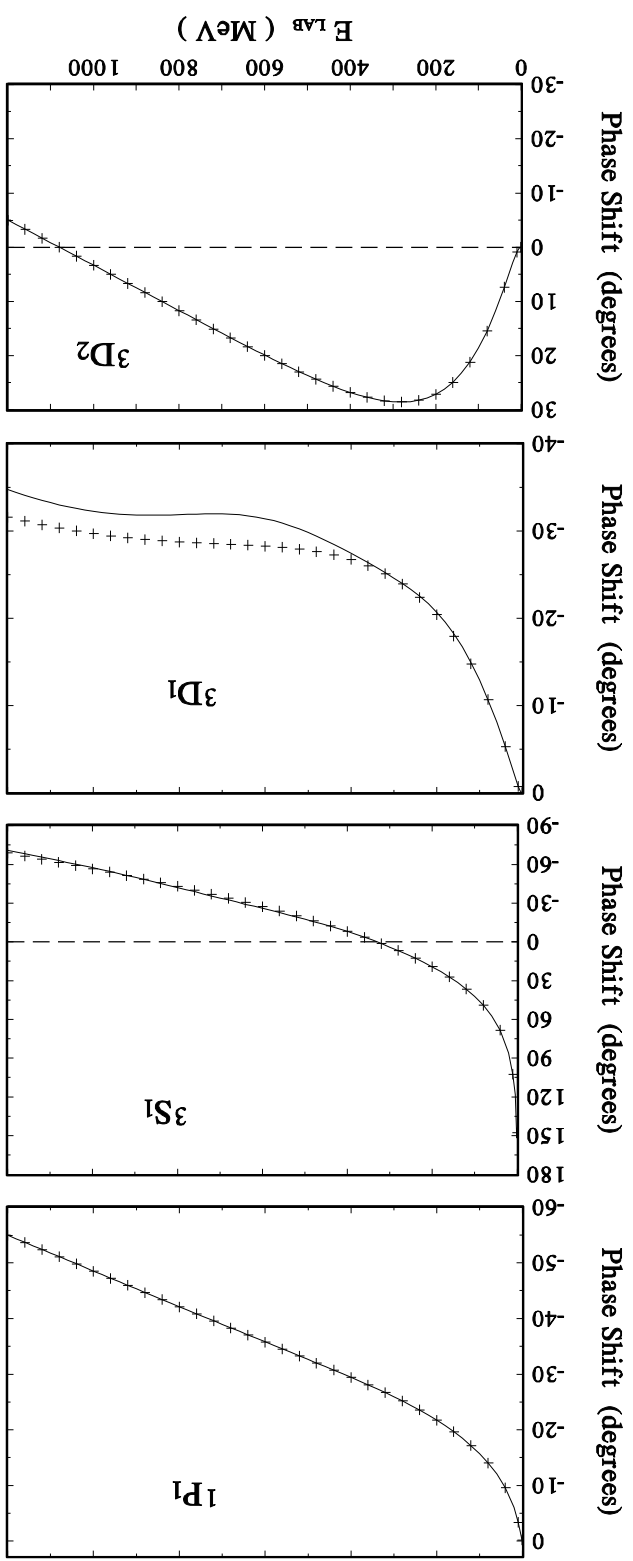
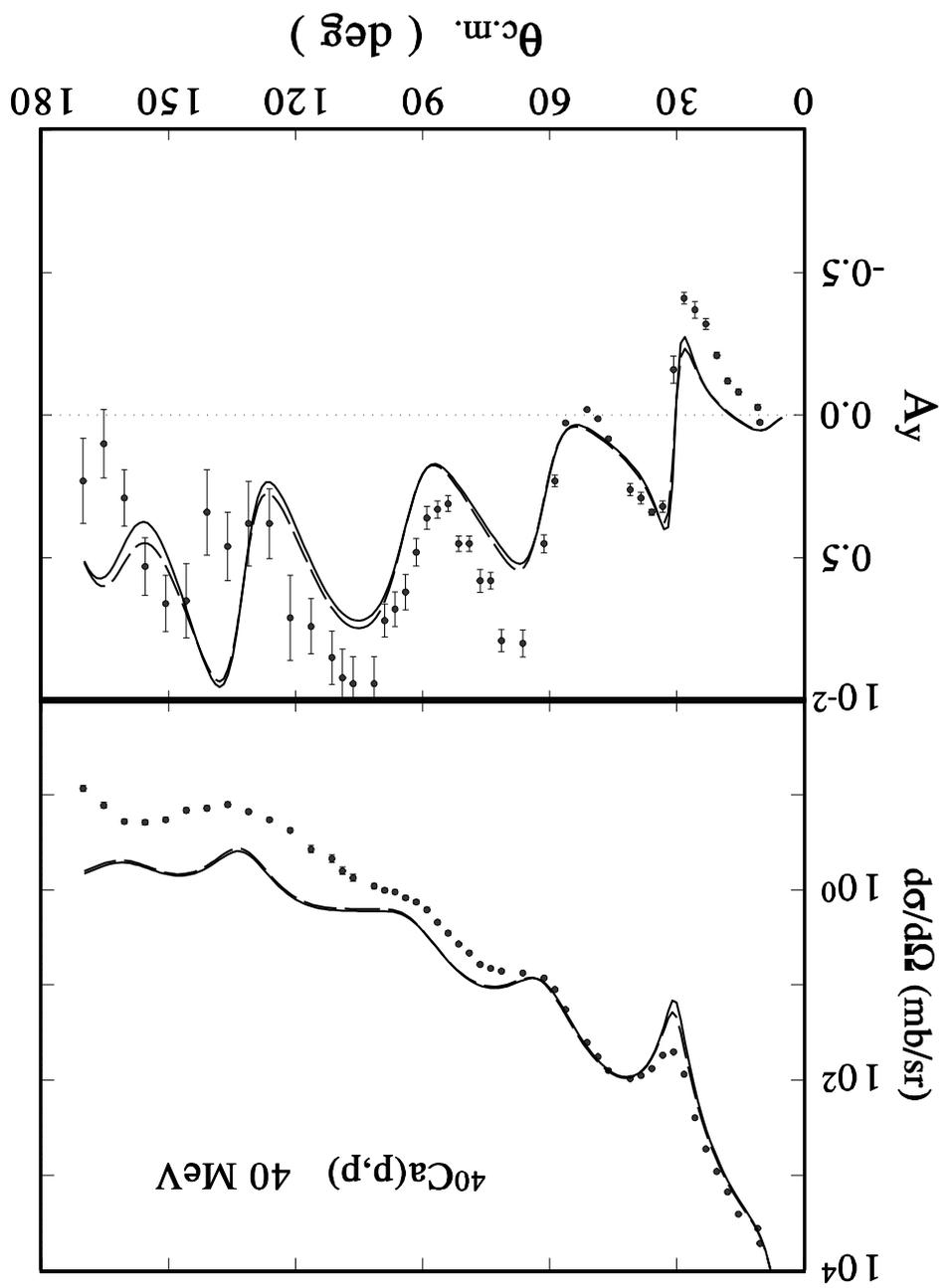
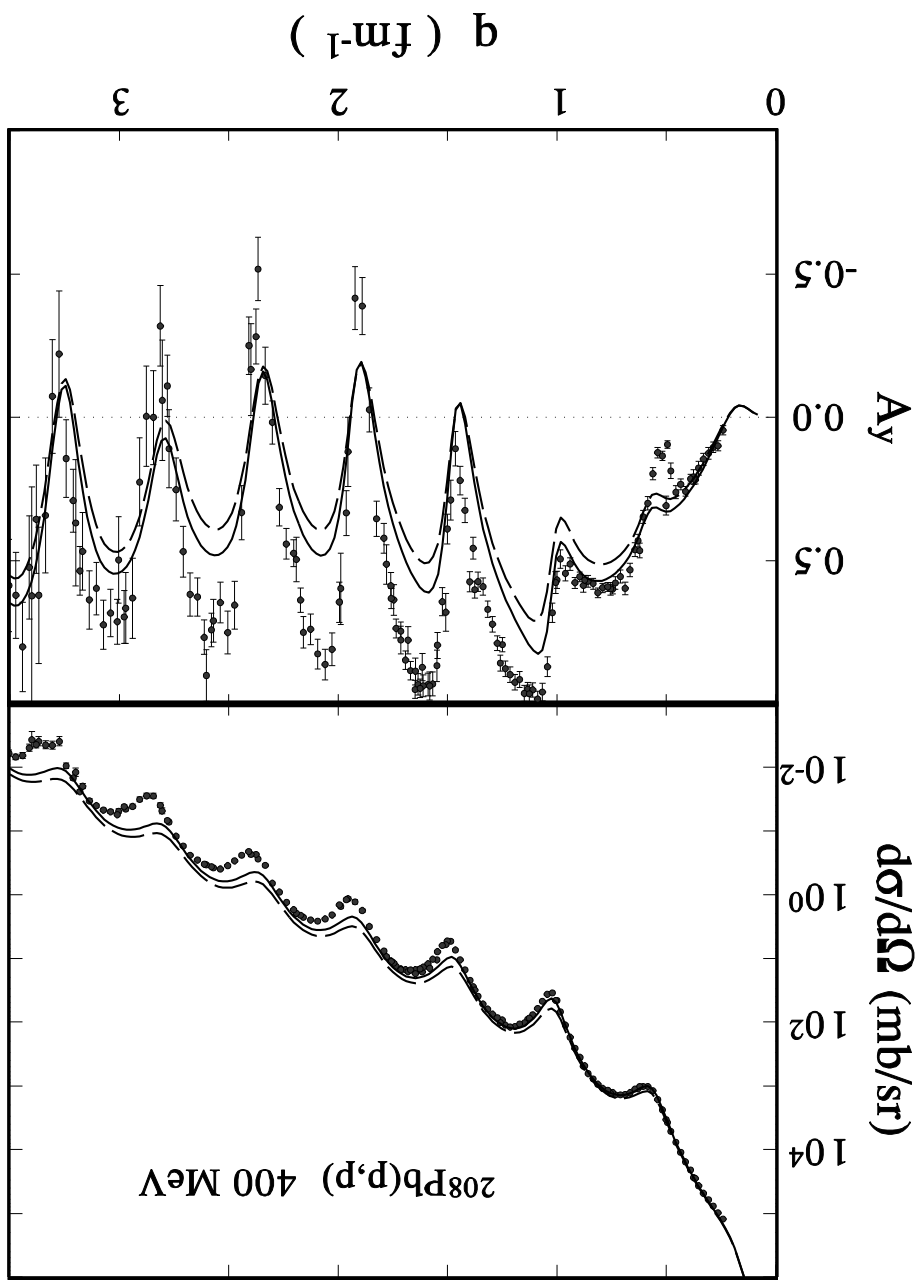
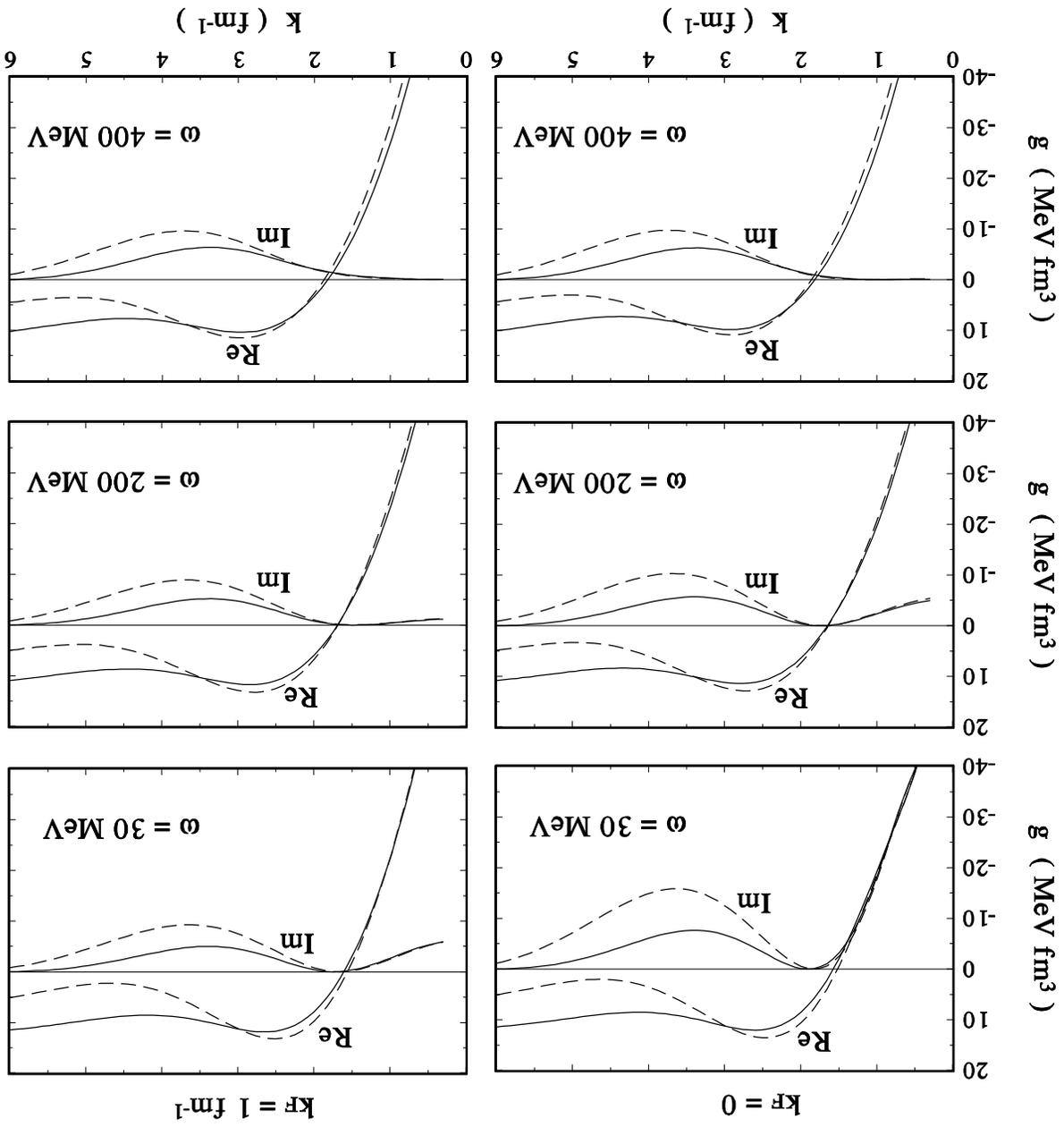
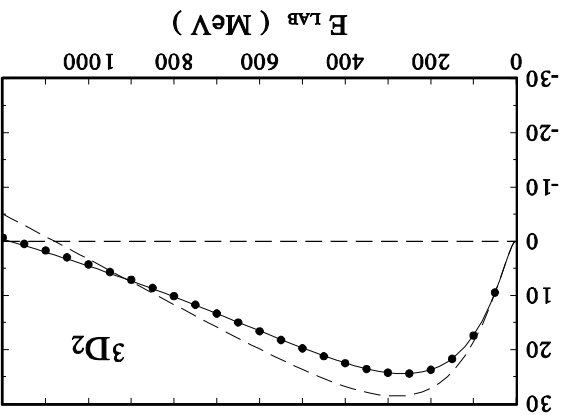
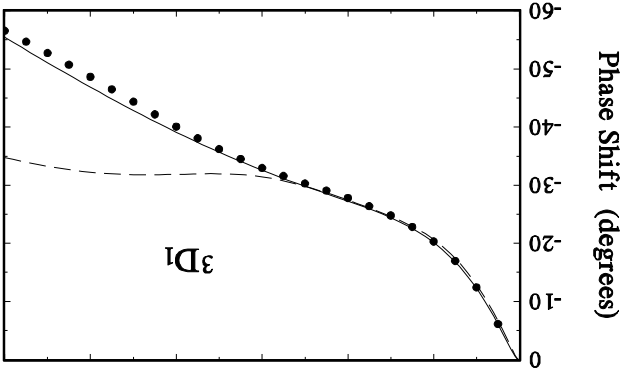
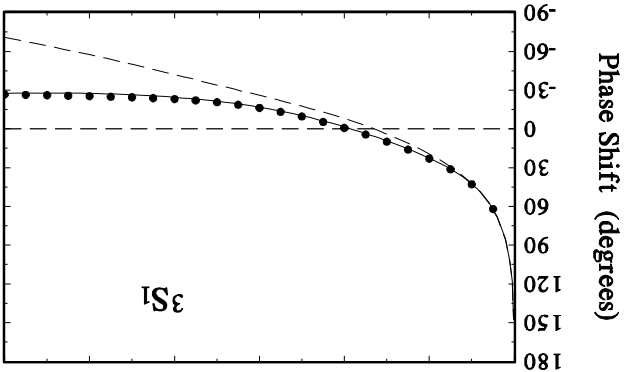
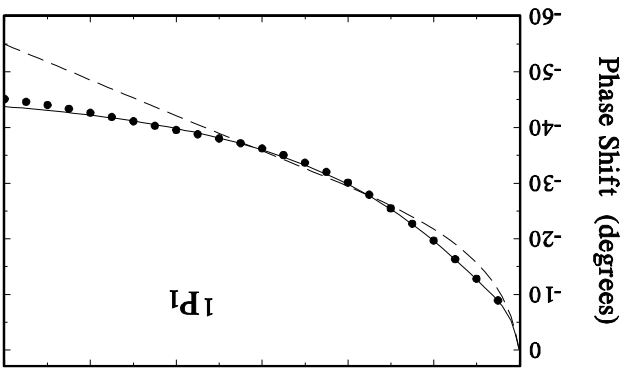
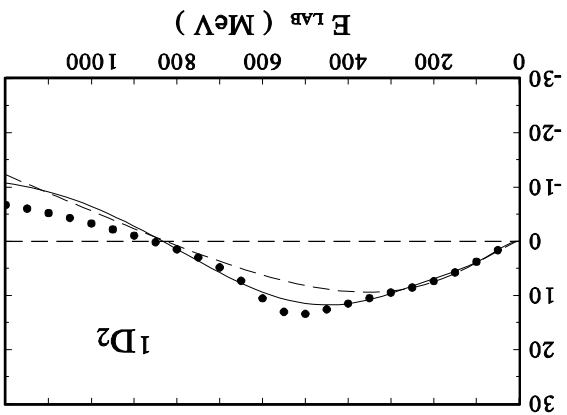
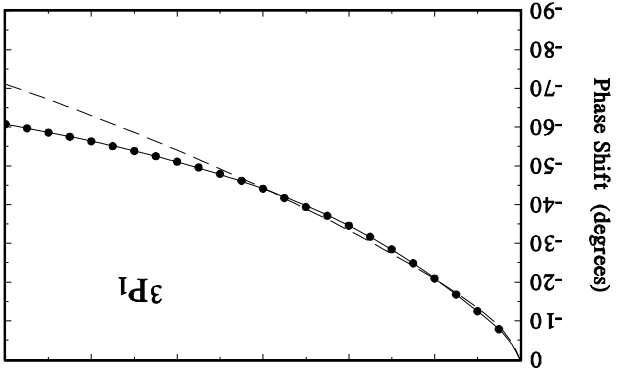
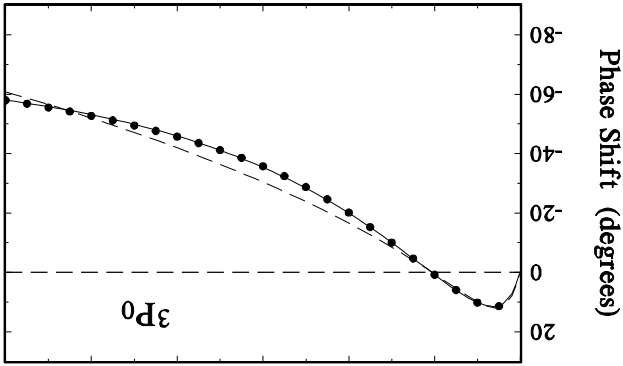
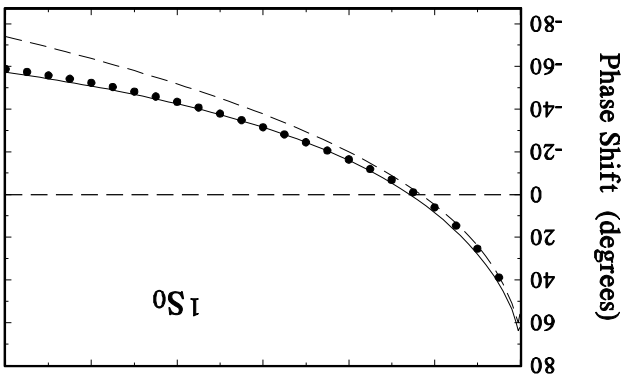
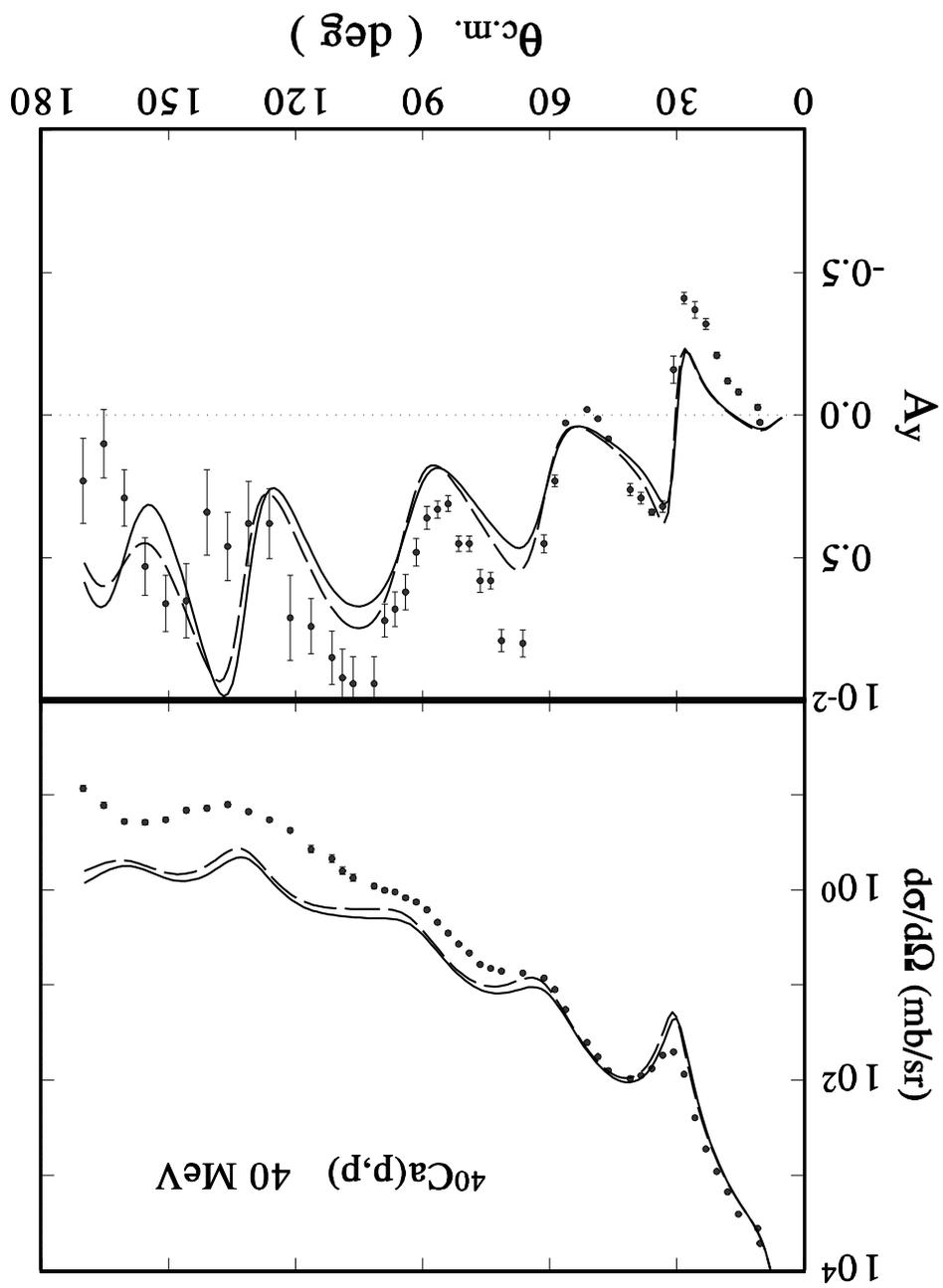
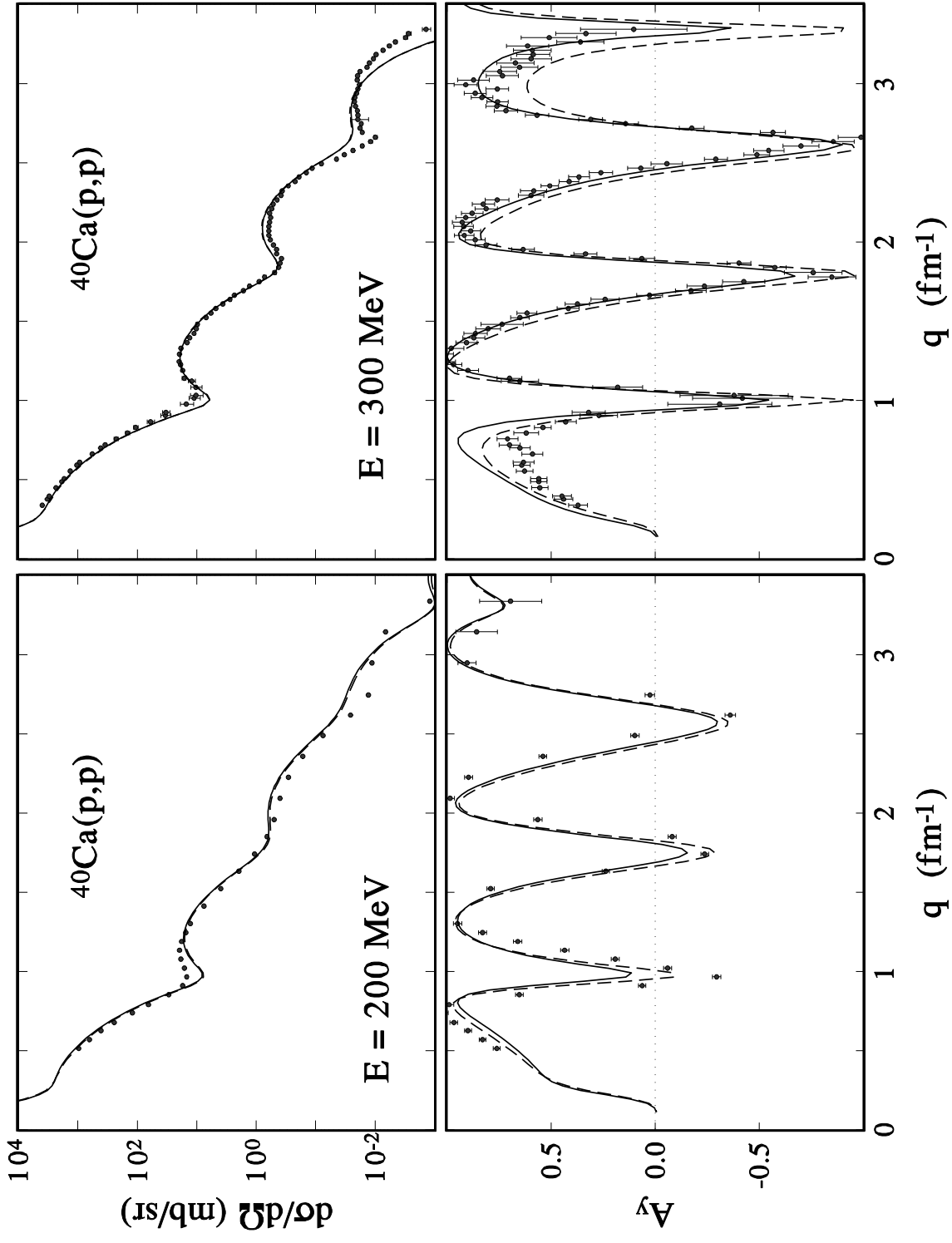
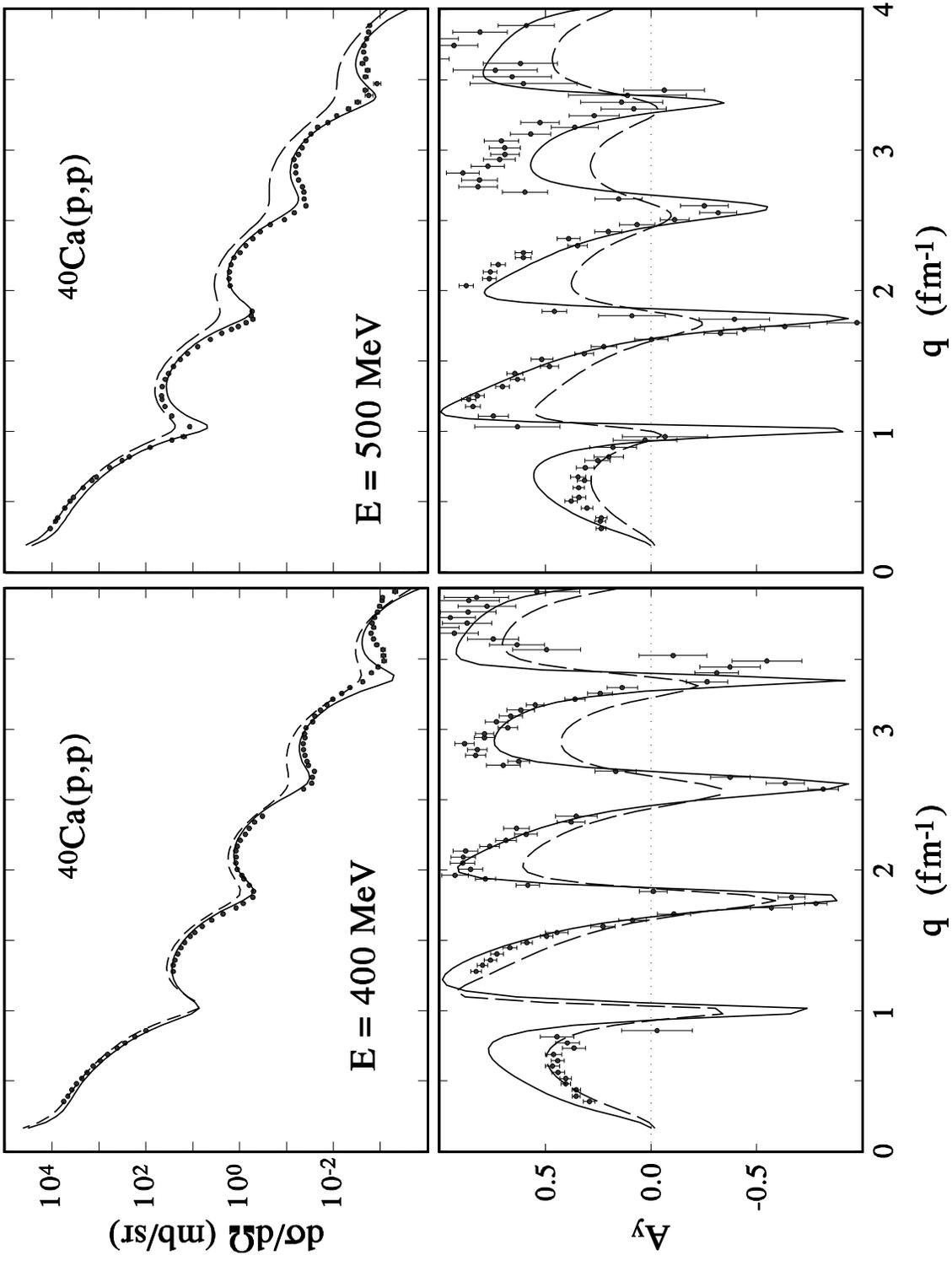
\end{document}